\documentclass[
superscriptaddress,
nofootinbib,
twocolumn,
 amsmath,amssymb,
 aps,
pra,
notitlepage,
longbibliography
]{revtex4-1}

\usepackage{dcolumn}
\usepackage{bm}
\usepackage{epsfig}
\usepackage{graphicx}
\usepackage{latexsym}
\usepackage{amsfonts}
\usepackage{setspace}
\usepackage{graphicx}
\usepackage{amsmath}
\usepackage{verbatim}
\usepackage{color}
\usepackage{SIunits}
\usepackage{hyperref}

\newcommand{\be}{\begin{equation}}
\newcommand{\ee}{\end{equation}}
\newcommand{\ba}{\begin{array}}
\newcommand{\ea}{\end{array}}
\newcommand{\bqa}{\begin{eqnarray}}
\newcommand{\eqa}{\end{eqnarray}}

\begin{document}

\title{High-efficiency fiber-to-chip interface for aluminum nitride quantum photonics}

\author{Mengdi Zhao} 
\affiliation{Holonyak Micro and Nanotechnology Laboratory and Department of Electrical and Computer Engineering, University of Illinois at Urbana-Champaign, Urbana, IL 61801 USA}
\affiliation{Illinois Quantum Information Science and Technology Center, University of Illinois at Urbana-Champaign, Urbana, IL 61801 USA}
\author{Woraprach Kusolthossakul} 
\affiliation{Holonyak Micro and Nanotechnology Laboratory and Department of Electrical and Computer Engineering, University of Illinois at Urbana-Champaign, Urbana, IL 61801 USA}
\affiliation{Illinois Quantum Information Science and Technology Center, University of Illinois at Urbana-Champaign, Urbana, IL 61801 USA}
\author{Kejie Fang} 
\email{kfang3@illinois.edu}
\affiliation{Holonyak Micro and Nanotechnology Laboratory and Department of Electrical and Computer Engineering, University of Illinois at Urbana-Champaign, Urbana, IL 61801 USA}
\affiliation{Illinois Quantum Information Science and Technology Center, University of Illinois at Urbana-Champaign, Urbana, IL 61801 USA}

\begin{abstract}
Integrated nonlinear photonic circuits received rapid development in recent years, providing all-optical functionalities enabled by cavity-enhanced photon-photon interaction for classical and quantum applications. A high-efficiency fiber-to-chip interface is key to the use of these integrated photonic circuits for quantum information tasks, as photon loss is a major source that weakens quantum protocols. Here, overcoming material and fabrication limitation of thin-film aluminum nitride by adopting a stepwise waveguiding scheme, we demonstrate low-loss adiabatic fiber-optic couplers in aluminum nitride films with a substantial thickness ($\sim$600 nm) for optimized nonlinear photon interaction.  For telecom (1550 nm) and near-visible (780 nm)  transverse magnetic-polarized light, the measured insertion loss of the fiber-optic coupler is -0.97 dB and -2.6 dB, respectively. Our results will facilitate the use of aluminum nitride integrated photonic circuits as efficient quantum resources for generation of entangled photons and squeezed light on microchips.
\end{abstract}

\maketitle

\section{Introduction}

Nonlinear optics is crucial for realizing key quantum optical resources, including heralded single photons \cite{hong1986experimental} and squeezed light \cite{slusher1985observation,wu1986generation}, which, however, have heavily relied on using bulk nonlinear crystals since the dawn of (experimental) quantum optics three decades ago. In recent years, development of wafer-scale growth of high-quality thin-film nonlinear optical materials and the associated microfabrication techniques is shifting the paradigm of quantum optics from the conventional bulk crystals and components to integrated photonic circuits on microchips, driven by the quest of scalable systems and strong light-matter interaction among other advantages of chip-scale architectures. Integrated nonlinear photonic circuits have now been developed in material systems including, for example, silicon \cite{silverstone2016silicon} and silicon nitride \cite{xiong2015compact,dutt2015chip,lu2019chip} with $\chi^{(3)}$ nonlinearity, and more recently aluminum nitride (AlN) \cite{xiong2012aluminum} and lithium niobate \cite{wang2019monolithic,luo2018highly} with dominant $\chi^{(2)}$ nonlinearity. The $\chi^{(2)}$ nonlinear materials are particularly useful for generating entangled photon pairs via spontaneous parametric down conversion \cite{guo2017parametric}, squeezed light \cite{lenzini2018integrated}, and visible-to-infrared wavelength conversion \cite{guo2016chip}. With centrosymmetry breaking of the crystal structure, these materials also possess large linear electro-optic coefficients, leading to applications such as integrated high-speed modulators \cite{wang2018integrated} and microwave-to-optical signal transducers for quantum networks \cite{fan2018superconducting}.

A key to the success of these applications of integrated nonlinear photonic circuits is a low-loss fiber-to-chip interface, which will help to preserve the fidelity of quantum states or to increase the success rate of post-selected events for enhanced quantum protocols. Typical fiber-to-chip coupling schemes have been realized in the AlN platform, including grating couplers \cite{lu2018aluminum,ghosh2012aluminum,pernice2012second} and end-fire couplers\cite{jung2014green,liu2017aluminum}. However, these coupling schemes generally have substantial insertion losses due to abrupt mode change, especially for the near-visible light (see Table \ref{tab:shape-functions} for a summary of demonstrated fiber-optic couplers in AlN). 

\begin{table*}[htbp]
\centering
\caption{\bf Fiber-to-chip coupling schemes realized in AlN thin films}
\begin{tabular}{c|c|c|c|c}
\hline
Coupling scheme & Reference & AlN film thickness & Wavelength (Polarization) & Single-side coupling efficiency \\
\hline
End-fire coupler & \cite{jung2014green} & 650 nm & 1550 nm (TE) & -4 dB   \\ \cline{2-5}
                            & \cite{liu2017aluminum} &1.2$\ \mu$m & 1550 nm (TM/TE) & -2.8 dB/-2.8 dB \\ \hline
Grating coupler & \cite{lu2018aluminum} & 200 nm & 500 nm-700 nm (TE)  & -4 dB \\ \cline{2-5}
                           & \cite{pernice2012second} & 330 nm & 780 nm (TM) & -11.5 dB  \\ 
  			  &&             & 1550 nm (TM) & -13.5 dB  \\ \cline{2-5}
  		 	  & \cite{ghosh2012aluminum} & 400 nm & 1550 nm (TE) & -6.6 dB  \\ \hline
Adiabatic coupler & this work & 600 nm & 780 nm (TM/TE)   & -2.6 dB/-2.6 dB  \\  \cline{4-5}
			 &		    &             & 1550 nm (TM/TE)   & -0.97 dB/-1.55 dB \\
\hline
\end{tabular}
  \label{tab:shape-functions}
\end{table*}

In order to exploit the largest $\chi^{(2)}$ component of $c-$plane AlN films, transverse magnetic (TM)-like modes of a photonic device should be used for both fundamental (1550 nm wavelength) and second-harmonic (780 nm wavelength) light \cite{guo2016chip}. In this work, integrated photonic circuits are realized in 600 nm thick AlN films; this thickness is chosen in order to maximize the nonlinear coupling between phase-matched fundamental TM$_{00}$ and second-harmonic TM$_{20}$ modes in a medium sized ring resonator ($r=30\ \mu$m), since the nonlinear mode coupling scales inversely with the square root of the ring radius \cite{guo2016second} and optical quality factor of resonances tends to reduce in smaller rings. However, 600 nm thick AlN films are not favorable for making either grating couplers or end-fire couplers: grating couplers are suitable for thinner AlN films due to the difficulty of fabricating gratings with high aspect ratio, especially for near-visible light, and end-fire couplers tend to work better with thicker AlN films where mode field diameter mismatch between waveguides and optical fibers decreases. As a result, a low-loss fiber-optic coupling scheme that works for AlN films with intermediate thickness (e.g. $\gtrsim300$ nm and $\lesssim1\ \mu$m), and especially for TM-polarized light, is needed.

Here, we realize a high-efficiency fiber-to-chip interface using adiabatic couplers in thin-film AlN. Adiabatic couplers, based on adiabatic band-crossing between a tapered optical fiber and a tapered on-chip waveguide to achieve near-unity transmission of light, have been demonstrated previously in material systems including silicon \cite{groblacher2013highly}, silicon nitride \cite{tiecke2015efficient}, and diamond \cite{burek2017fiber}. Due to the specific material property and fabrication difficulty associated with AlN, as we will detail below, realizing low-loss adiabatic couplers in this material represents a substantial challenge, especially for the near-visible TM-polarized light. By adopting a stepwise waveguiding scheme, we overcome these limitations and realize adiabatic couplers for both 780 nm and 1550 nm TM-polarized light, with an insertion loss of -2.6 dB and -0.97 dB, respectively. It turns out the transmission of the transverse electric (TE)-polarized light is also high in the same couplers that are optimized for the TM polarization.

\section{Simulation results}

\begin{figure}[!htb]
\begin{center}
\includegraphics[width=1\columnwidth]{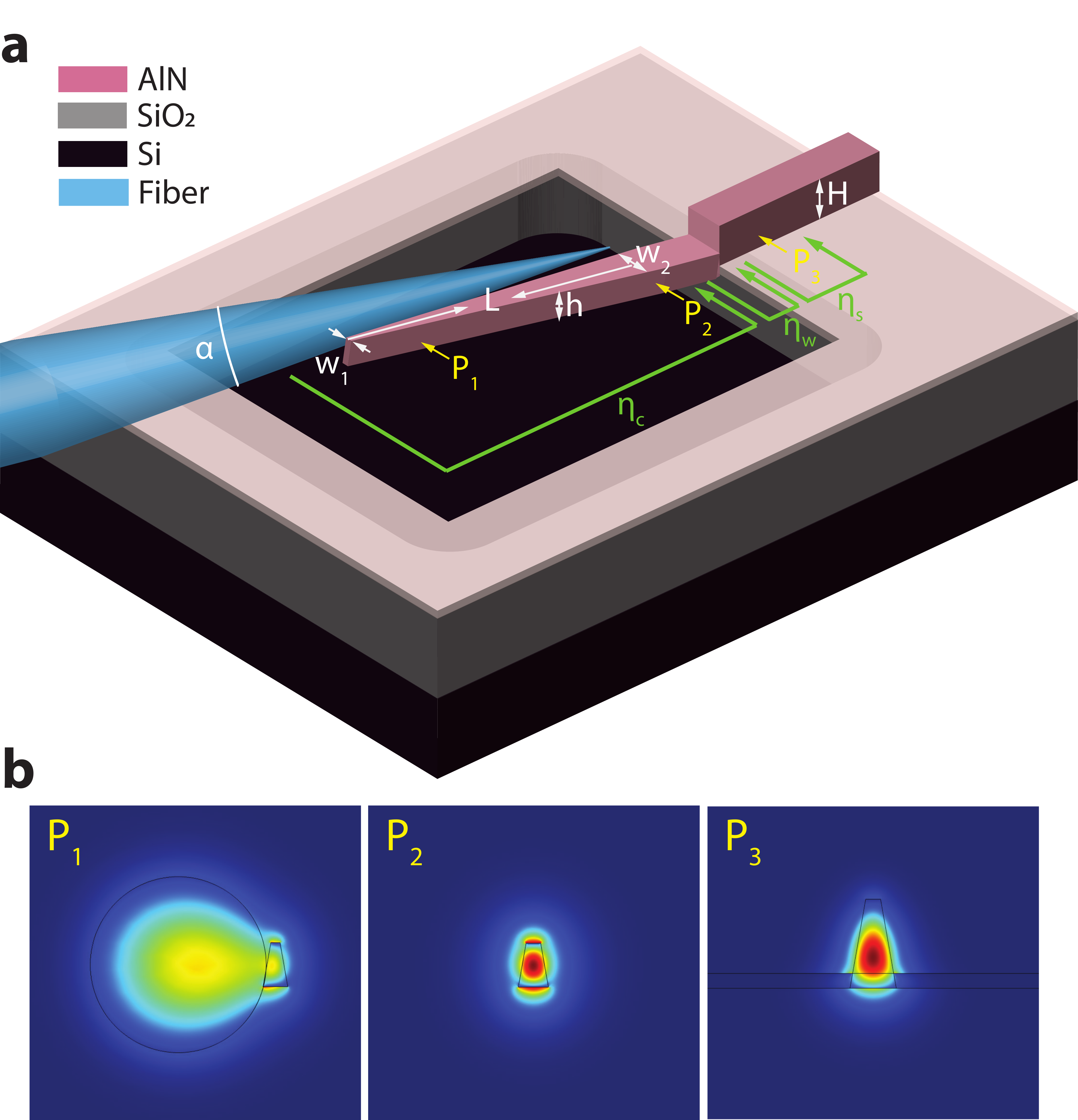}
\caption{\textbf{a}. Schematic diagram of the adiabatic fiber-optic coupler connected with stepwise waveguide in AlN-on-insulator microchips. The tapered optical fiber is side-coupled to the tapered on-chip waveguide coupler. \textbf{b}. $E_z$ field of the 773 nm wavelength fundamental TM-like mode of the coupled fiber-waveguide system at locations $P_{1,2,3}$. $\eta_{c,w,s}$ are the transmission efficiency of different sections of the coupler, explained in the text.} 
\label{fig1}
\end{center}
\end{figure}

As mentioned above, specific constraints due to the material properties of AlN and the extreme dimensions of adiabatic couplers needed for high transmission efficiency make the realization of adiabatic couplers in relatively thick AlN films an outstanding challenge, especially for the 780 nm wavelength light. High-efficiency adiabatic couplers require the tip of the tapered waveguide to be sufficiently small to avoid insertion loss when interfacing with the low-index optical fiber. This is especially critical when dealing with TM-like modes, which have larger index than TE-like modes in waveguides with large height-to-width ratio. In contrast, most demonstrated couplers in AlN films with reasonable insertion loss work for TE-like modes in the telecom wavelength range (Table \ref{tab:shape-functions}). However, in practice, the lateral dimension of lithographically fabricated AlN waveguides is limited by the lithography resolution and the tapering-out effect due to the dry etching process with best sidewall angles around 80$\degree$ achieved so far\cite{guo2016chip,sohn2018time}. For example, in this work we used hydrogen silsesquioxane (HSQ) as the electron beam resist, and the smallest feature size that can be patterned on a 300 nm thick HSQ resist without mask collapse after development is about 50 nm. Given the etch-defined sidewall angle of AlN waveguides to be 80$\degree$, the smallest achievable dimension of the waveguide bottom is around 260 nm assuming a 600 nm thick AlN film. The 780 nm fundamental TM-like mode in such waveguides has a mode index of 1.65, which is larger than the index of the TM mode of a bare optical fiber, indicating substantial insertion loss for the adiabatic coupler. The adiabatic couplers realized based on such dimensions show an efficiency of only about 1$\%$ according to our simulation.

To overcome these challenges, we developed a stepwise coupling scheme for the 780 nm wavelength light. By selectively thinning the thickness of the adiabatic coupler, this scheme  simultaneously realizes strong nonlinear mode coupling in photonic devices with the full thickness of AlN films and efficient fiber-to-chip transduction via the thinned adiabatic coupler. A schematic of the stepwise coupler in AlN-on-insulator microchips is shown in Fig. \ref{fig1}a. The thickness of the AlN film is $H$. The released AlN waveguide coupler, anchored at an unreleased rib waveguide, has a thickness of $h$ and is adiabatically tapered from a width of $w_1$ to $w_2$ over a length of $L$. A tapered optical fiber is aligned side to the released waveguide over the tapered region to form the adiabatic fiber-optic coupler. We emphasize the  side-coupling of fiber and waveguide is important here as top-coupling results in much weaker interaction between TM-like modes of the AlN waveguide and the optical fiber in thick AlN films. As the simulation shows, for the 780 nm wavelength light, the released waveguide with reduced thickness (i.e. $h<H$) significantly increases the efficiency of the adiabatic coupler for the fundamental TM-like mode. Though the step between the released coupler and the rib waveguide introduces some loss, the overall transmission efficiency from the fiber to the on-chip waveguide, for the optimal $h$, is much higher than an adiabatic coupler with full thickness $H$.  Fig. \ref{fig1}b shows the 780 nm fundamental TM-like mode of the coupled fiber-waveguide system at a few locations for the optimized case, simulated using COMSOL.

\begin{figure}[!htb]
\begin{center}
\includegraphics[width=1\columnwidth]{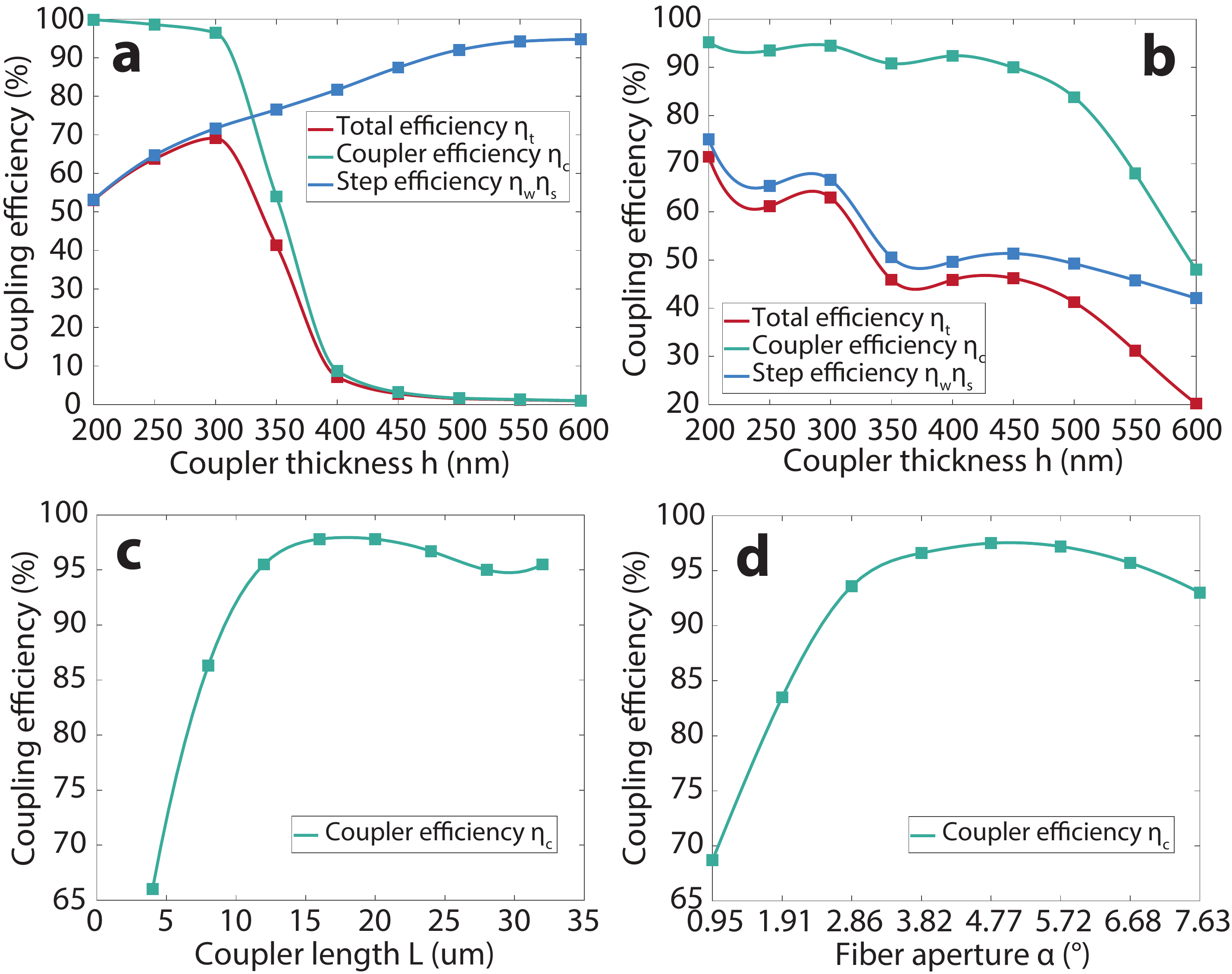}
\caption{\textbf{a}. Optimized coupling efficiency for 773 nm wavelength TM polarized light for various thickness of the tapered waveguide. Red is the optimized overall coupling efficiency ($\eta_t=\eta_c\eta_w\eta_s$), and green and blue are the corresponding efficiency of the adiabatic coupler $\eta_c$ and the step $\eta_w\eta_s$, respectively. \textbf{b}. Coupling efficiency of the TE-polarized light through the same structure that optimizes the TM polarization in \textbf{a}. \textbf{c}. Adiabatic coupler efficiency $\eta_c$ versus tapering length $L$ for $\alpha=3.8\degree$, $h=300$ nm, $w_1=102$ nm, and $w_2=150$ nm. \textbf{d}. Adiabatic coupler efficiency $\eta_c$ versus fiber aperture $\alpha$ for $L=24\ \mu$m, $h=300$ nm, $w_1=102$ nm, and $w_2=150$.} 
\label{fig2}
\end{center}
\end{figure}

We used Lumerical FDTD for the simulation of coupler efficiency. In the simulated structures, we assume $H=600$ nm, and the rib waveguide is 500 nm tall above the 100 nm thick AlN membrane. The cross-section of the AlN coupler is set to be a rectangle with its width equal to the medium width of the actual trapezoidal waveguide cross-section, i.e. $\bar w=w+h/\tan(80\degree)$, where $w$ is the width of the trapezoidal waveguide top and we assumed the dry-etch induced sidewall angle to be $80\degree$. This is a legitimate approximation of the actual trapezoidal waveguide cross-section, since for adiabatic couplers, the coupling efficiency is insensitive to the shape of the waveguide cross-section but only depends on the adiabatic varying of the mode index and fiber-waveguide mode coupling strength. As long as the mode index is close, adiabatic couplers with different cross-sections will yield similar transmission efficiency. For example, for the 780 nm TM$_{00}$ mode, its mode index is 1.125 and 1.134, respectively, for a trapezoidal waveguide with 50 nm wide top, 300 nm thickness and 80$\degree$ base angle and a rectangular waveguide with 102 nm width (i.e. the medium width of the trapezoid) and the same thickness. From here on, the width of a simulated waveguide thus refers to the width of the equivalent rectangular waveguide.

We first fix $L=24\ \mu$m, $w_1=102$ nm, and the aperture of tapered fibers $\alpha=3.8\degree$. We used refractive index 2.183 and 2.140 for the 780 nm TM and TE modes, respectively. For the 780 nm wavelength TM$_{00}$ coupler, we varied $h$ and for each $h$ swept $w_2$ to maximize the adiabatic coupler efficiency $\eta_c$. The total coupling efficiency from the optical fiber to the unreleased rib waveguide is calculated as $\eta_t=\eta_c\eta_w\eta_s$, where $\eta_w$ and $\eta_s$ are the transmission efficiency at the waveguide step and the substrate etch front, respectively. Using this protocol, Fig. \ref{fig2}a shows the simulated $\eta_t$ versus $h$ for TM$_{00}$ mode and wavelength $\lambda=773$ nm. We find $h=300$ nm is optimal for TM$_{00}$ mode, with $\eta_t=69\%$ (wherein $\eta_c=97\%$, $\eta_w=75\%$, $\eta_s=95\%$, respectively) achieved at $w_2=150$ nm. Fig. \ref{fig2}b shows the coupling efficiency for TE$_{00}$ mode versus $h$ in the same structure that optimizes the coupling efficiency of TM$_{00}$ mode. We find the coupling efficiency for the TE$_{00}$ mode is $63\%$ for $h=300$ nm. We stress that this is not the optimized result for the TE$_{00}$ mode which is not the focus of this work. 
To justify the choice of taper length and fiber aperture used above, we simulated the adiabatic coupler efficiency $\eta_c$ as a function of the coupler length $L$ (with fixed fiber aperture $\alpha=3.8\degree$) and fiber aperture (with fixed coupler length $L=24\ \mu$m), respectively, with all the other parameters kept the same as the structure that optimized the total coupling efficiency, i.e. $h=300$ nm, $w_1=102$ nm, $w_2=150$ nm. The results are shown in Fig. \ref{fig2}c and d. It can be seen that both $L=24\ \mu$m and $\alpha=3.8\degree$ saturate the coupling efficiency, thus validating their choice.

\begin{figure}[!htb]
\begin{center}
\includegraphics[width=1\columnwidth]{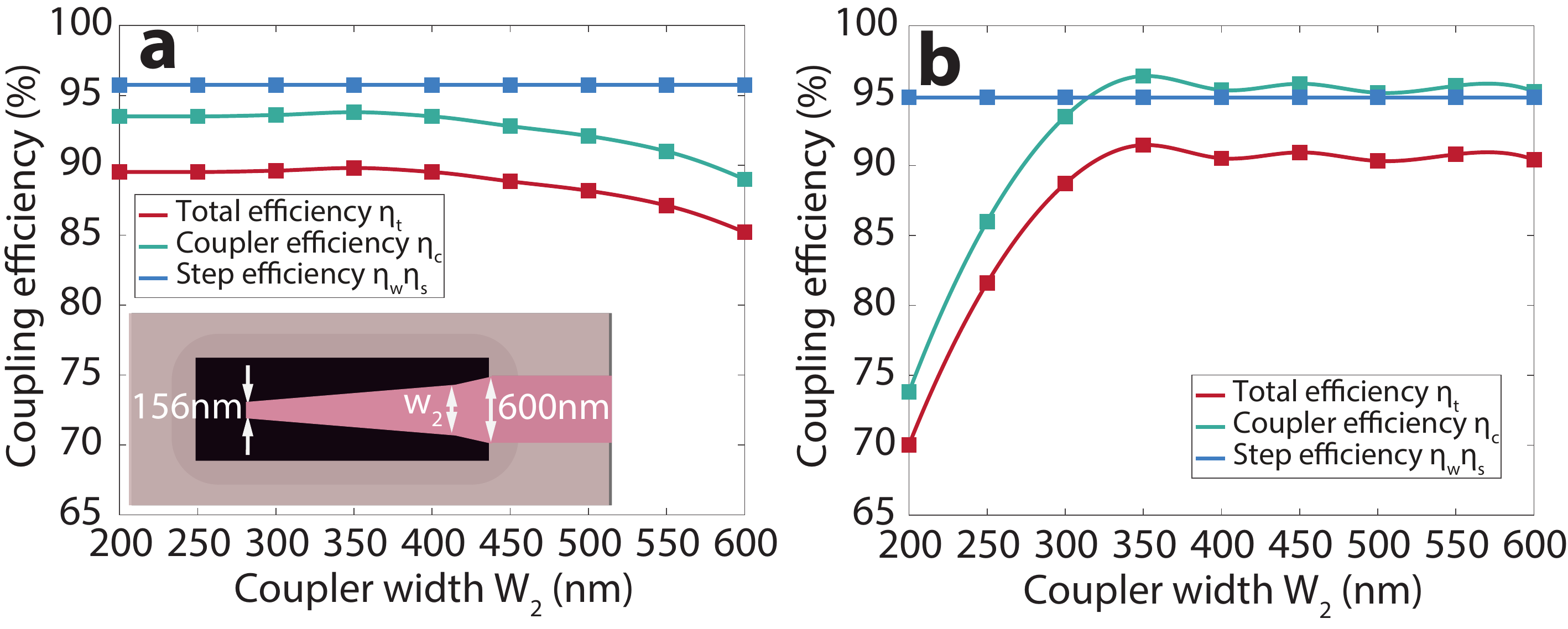}
\caption{\textbf{a}. Optimized coupling efficiency for 1546 nm wavelength TM polarized light for various $w_2$. Red is the optimized overall coupling efficiency, and green and blue are the corresponding efficiency of the adiabatic coupler and the waveguide step, respectively. Inset: dimensions of the coupler with an additional 5 $\mu$m long tapering from $w_2$ to 600 nm wide rib waveguide. \textbf{b}. Coupling efficiency of the TE-polarized light through the same structure that optimizes the TM polarization in \textbf{a}. } 
\label{fig3}
\end{center}
\end{figure}

In contrast, for the TM$_{00}$ mode at telecom wavelength, the released adiabatic coupler with thickness $h=H$ is able to achieve near-unity efficiency as the mode index at the waveguide tip with practical dimensions is still sufficiently smaller than that of the optical fiber mode. Thus, we adopt a flattop coupler design for the telecom wavelength light. In the simulation, we used refractive index 2.16 and 2.12 for the 1550 nm TM and TE modes, respectively, and we again used a rectangular waveguide cross-section for simplicity; this is justified because, for the 1550 nm TM$_{00}$ mode, its mode index is very close, i.e. 1.054 and 1.063 respectively, in a trapezoidal waveguide with 50 nm wide top, 600 nm height, and 80$\degree$ base angle and a rectangular waveguide with 156 nm width (i.e. the medium width of the trapezoidal waveguide) and 600 nm height. The design is straightforward: we fixed $L=24\ \mu$m, $w_1=156$ nm, $\alpha=3.8\degree$, and varied $w_2$ to optimize the coupling efficiency for TM$_{00}$ mode at $\lambda=1546$ nm. The width of the rib waveguide is kept at 600 nm so there is a second taper from $w_2$ to 600 nm over $5\ \mu$m length (Fig. \ref{fig3}a inset). Since the light has already been confined in the waveguide at the width $w_2$, this second taper has negligible transmission loss. Here, the step efficiency $\eta_w$ is merely due to the existence of the 100 nm AlN membrane. The simulated result is shown in Fig. \ref{fig3}a and the optimal coupling efficiency is found to be $91\%$ for $w_2=350$ nm. The coupling efficiency of the TE$_{00}$ mode is also simulated for the same structure, as shown in Fig. \ref{fig3}b, and for $w_2=350$ nm it is $91\%$. 

\section{Fabrication and measurements} 

We fabricated the designed couplers in AlN-on-oxide microchips with 600 nm thick AlN films deposited via a sputtering process with dual cathode S-gun magnetron source. The process is tuned such that the AlN thin-film is deposited with 400 MPa tensile stress. Due to this tensile stress, the released couplers bend up out of the plane of chip, enabling alignment and side-coupling with tilted optical fibers (to avoid touching the chip). The integrated AlN photonic circuits with released adiabatic couplers for both telecom and near-visible wavelength light are fabricated using the following process. First, a window enclosing the to-be-fabricated 780 nm wavelength coupler is lithographically patterned followed by removal of 300 nm AlN thereof using Cl$_2$/BCl$_3$/Ar-based inductively-coupled plasma-reactive ion etch (ICP-RIE). Etching down the window prior to patterning the 780 nm wavelength coupler ensures the width of the coupler top close to the lithography resolution and be consistent with the simulation setting. Then the whole photonic circuits including the adiabatic couplers are patterned using electron beam lithography with HSQ mask followed with ICP-RIE to etch away 500 nm AlN. Finally, a window enclosing the 1550 nm wavelength coupler, which is still protected by the residual HSQ from the previous beam write, is patterned and etched down by about 100 nm. At this point, both windows enclosing the adiabatic couplers are etched through while the rest photonic circuit is attached to the remaining 100 nm thick AlN layer on the oxide substrate. Finally, the couplers are released using hydrofluoric acid (HF) etch. On the other hand, the tapered fibers are fabricated using HF etch \cite{turner1984etch,hong2014nanoscale}, by slowly pulling the fiber out of the HF liquid with controlled speed. The fabricated fiber has an aperture $\sim 4\degree$.

\begin{figure}[!htb]
\begin{center}
\includegraphics[width=0.8\columnwidth]{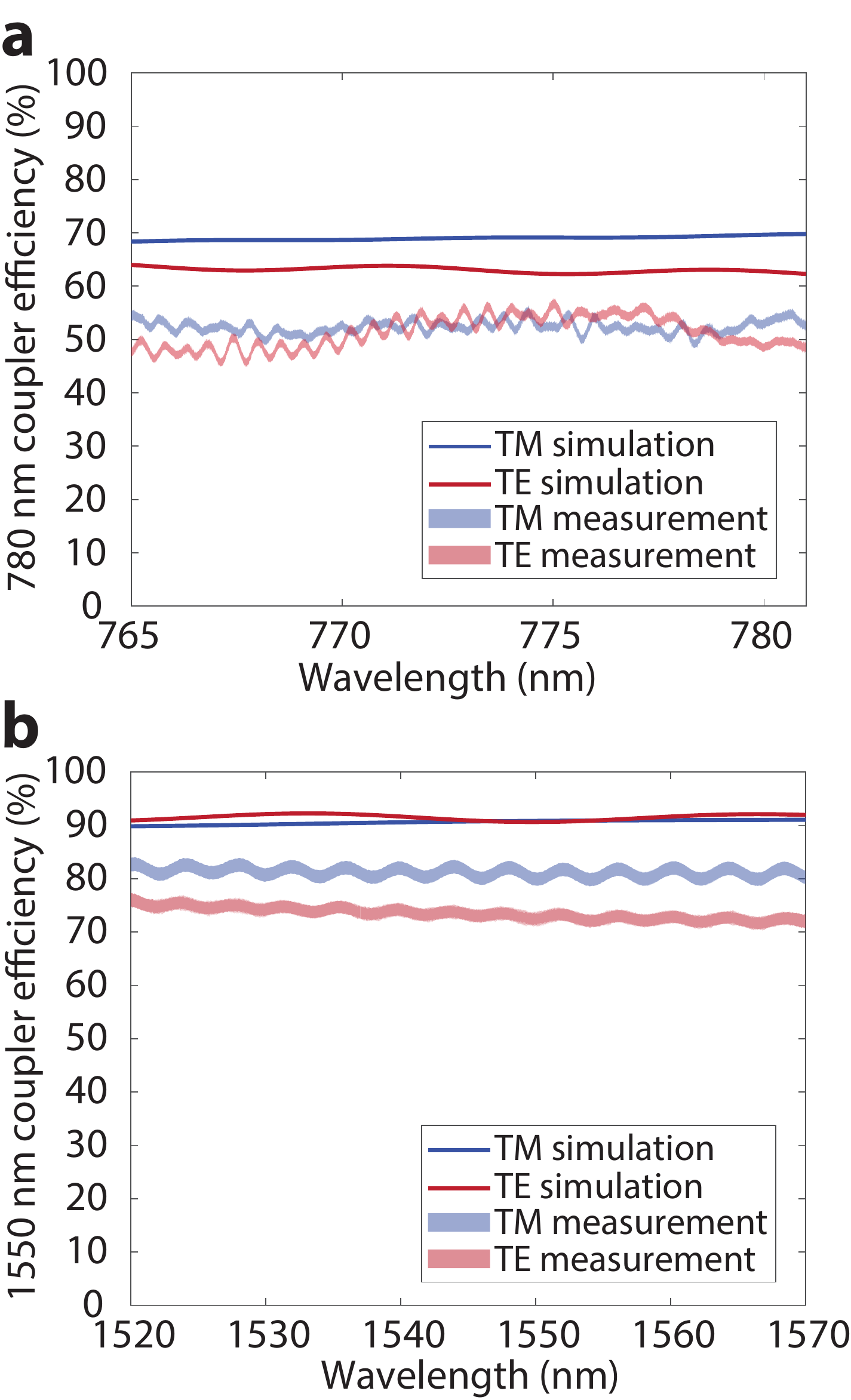}
\caption{\textbf{a} and \textbf{b}. Measured coupling efficiency for near-visible (\textbf{a}) and telecom (\textbf{b}) wavelength light in comparison with the simulation. Measured data is illustrated by counting for the variation due to fiber splicing loss which is about $3\%$ on average.} 
\label{fig4}
\end{center}
\end{figure}

\begin{figure*}[!htb]
\begin{center}
\includegraphics[width=2\columnwidth]{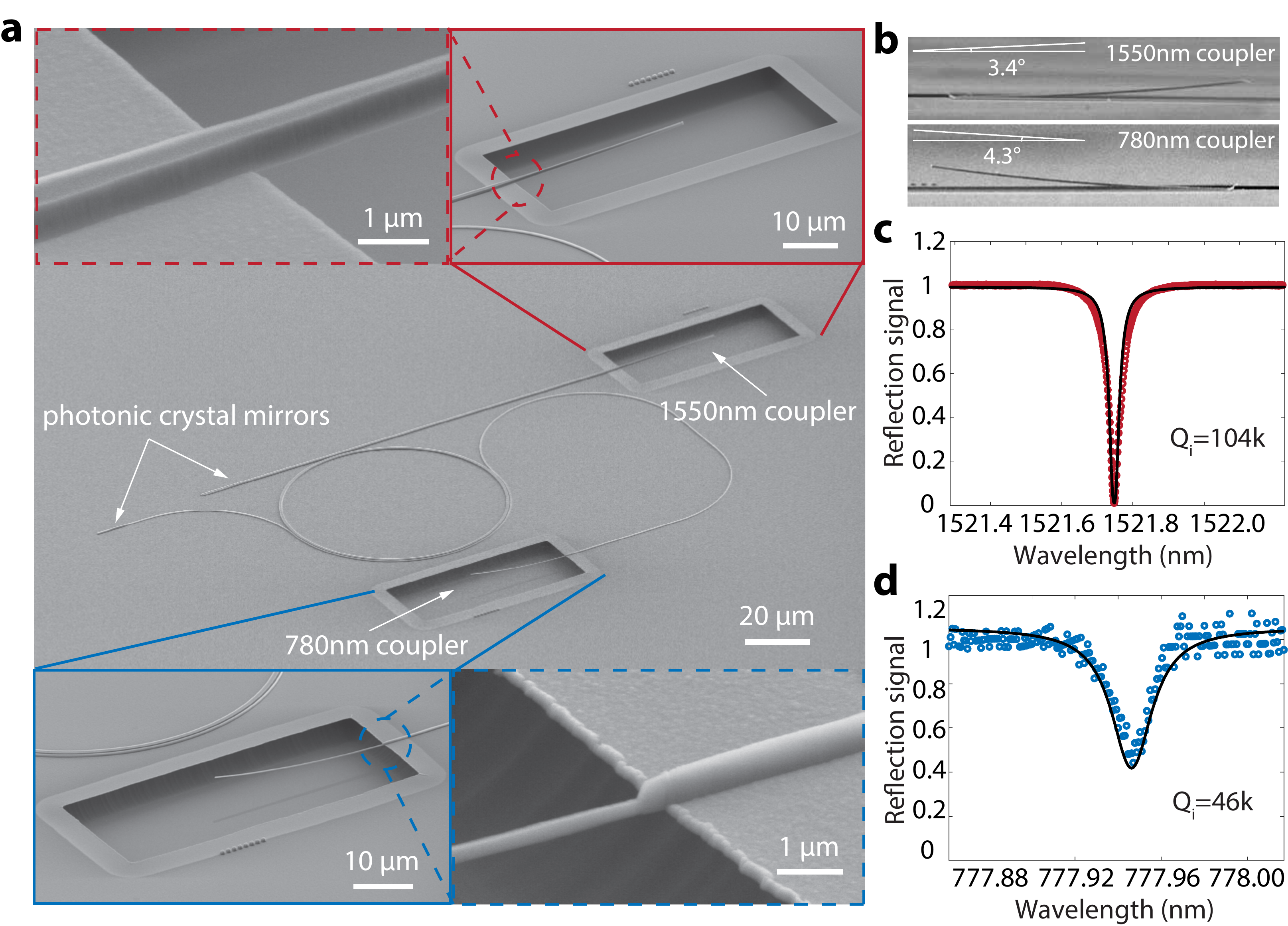}
\caption{\textbf{a}. SEM images of the AlN integrated photonic circuit with a bus waveguide-coupled ring resonator and both 780 nm and 1550 nm wavelength adiabatic couplers. \textbf{b}. Side-view SEM images of the 1550 nm wavelength coupler and 780 nm wavelength coupler. \textbf{c} and \textbf{d}. Normalized reflection signal showing a 1550 nm TM$_{00}$ resonance (\textbf{c}) and a 780 nm TM$_{20}$ resonance (\textbf{d}) of the ring resonator, respectively.} 
\label{fig5}
\end{center}
\end{figure*}
To test the efficiency of fabricated couplers, we made a photonic circuit consisting of two couplers connected by a short rib waveguide $40\ \mu$m long and aligned them with two tapered fibers, respectively. A wavelength-tunable diode laser is used to provide a few mW power of light to the input fiber and the transmission through the circuit is measured at the output fiber using a photodetector. The tapered fibers are mounted on adjustable holders tilted at an angle $\sim 4\degree$ relative to the plane of chip. The position of the fibers can be adjusted precisely using a set of motorized stages. The alignment of the fiber and released waveguide is achieved combining visual inspection of the fiber via an optical microscope and real-time monitoring of the optical transmission. The average efficiency of the two couplers is characterized by $\bar\eta_t=\sqrt{T}$, where $T$ is the fiber-to-fiber transmission efficiency. We find the result is consistent for repeated alignment and measurement on a same device and also across several devices with the same structure. Fig. \ref{fig4} shows the measured $\bar\eta_t$ of one device for the TM and TE polarized light in the 780 nm and 1550 nm bands.  The coupling efficiency for the 780 nm wavelength light is higher than $55\%$ (i.e. insertion loss of -2.6 dB) for both polarizations, and for the 1550 nm wavelength light it is over $80\%$ (-0.97 dB) for the TM polarization and over $70\%$ (-1.55 dB) for the TE polarization, respectively. One possible reason for the discrepancy between the measured coupling efficiency and the simulation is due to the curled shape of the suspended waveguide (Fig. \ref{fig5}b), which is different from the straight coupler used in the simulation. This could be ameliorated by using shorter couplers, for example, a 16 um long coupler which has similar efficiency as the 24 um long one (Fig. \ref{fig2}c), or only using the section from the tip to the middle of the released waveguide for tapering where it is quite straight.

The demonstrated low-loss fiber-optic couplers are ready to interface with various integrated photonic circuits for exploiting the optical nonlinearity of AlN. As an example, we designed and fabricated a photonic circuit with a ring resonator coupled with both 780 nm and 1550 nm wavelength bus waveguides. Such devices can be used for efficient generation of entangled photon pairs and squeezed light on chips. A scanning electron microscopy (SEM) image of the full device is shown in Fig. \ref{fig5}a. The 780 nm wavelength bus waveguide with a width of 80 nm, curled around the ring resonator to increase the coupling efficiency with the latter, is designed such that its TM$_{00}$ mode is phase-matched with the TM$_{20}$ mode of the ring resonator. The ring resonator itself with 30 $\mu$m radius and 740 nm width is designed so that its 780 nm wavelength TM$_{20}$ mode is phase-matched with the 1550 nm wavelength TM$_{00}$ mode for double-resonance enhanced nonlinear photon interaction.  Both waveguides are terminated with photonic crystal mirrors for measurement of reflected light using a single fiber. Fig.\ref{fig5}c and d show measured TM$_{00}$ and TM$_{20}$ resonances of a ring resonator, with an intrinsic quality factor of about $10^5$ and $0.5\times 10^5$, respectively. 

\section{Summary}
In summary, we have demonstrated a highly efficient fiber-to-chip interface in thin-film AlN using adiabatic fiber-optic couplers together with stepwise waveguiding. The measured insertion loss is as low as -0.97 dB and -2.6 dB for the 1550 nm and 780 nm wavelength fundamental TM-like modes, respectively, which provides a solution for low-loss fiber-optic coupling in AlN films with a thickness $\gtrsim300$ nm and $\lesssim1\ \mu$m where grating couplers and end-fire couplers are considerably lossy. The coupling efficiency of the fiber-optic couplers demonstrated here can be further improved by, for example, tuning ICP-RIE process to make narrower waveguides and adjusting the design to take into account of the morphology of the released waveguide. Since for most quantum photonic applications exploiting parametric down conversion, 780 nm wavelength light merely acts as the classical pump while 1550 nm wavelength light contains the useful non-classical correlations, the realized fiber-optic couplers here ensure a strong on-chip pump while minimize the transduction loss of quantum resources. This work will help establishing chipscale entangled photon pairs and squeezed light sources in thin-film AlN with strong $\chi^{(2)}$ nonlinearity.

\bigskip
\noindent\textbf{Funding.} National Science Foundation (NSF) (DMS 18-39177).

\bigskip
\noindent\textbf{Disclosures.} The authors declare no conflicts of interest.

%
 
\end{document}